\documentclass[seceqn]{elsart}
\usepackage{amsfonts}
\usepackage{amsmath}
\usepackage{amssymb}
\usepackage{graphicx}
\usepackage{subfigure}
\usepackage[square,comma,numbers,sort&compress]{natbib}
\usepackage{dsfont}
\usepackage{courier}
\usepackage{hyperref}

\input{tcilatex}
\renewcommand{\mathbb}[1]{{\mathds #1\/}}

\renewcommand{\newline}{\\}

\begin{document}

\begin{frontmatter}

\title{The Effects of Lagrangian vs. Population Dynamics
on Plankton Patchiness Generation}%\vspace{-1cm}
\author{M.~J. Olascoaga\corauthref{cor1}\thanksref{label1}},
\ead{jolascoa@rsmas.miami.edu} \corauth[cor1]{Corresponding author.
Tel.: +1-305-421-4873; fax: +1-305-421-4701.}
\author{F.~J. Beron-Vera\thanksref{label1}},
\author{M.~G. Brown\thanksref{label1}},
\author{H. Ko\c{c}ak\thanksref{label2}}
\address[label1]{RSMAS, University of Miami, Miami, FL 33149, USA}
\address[label2]{Departments of Computer Science and Mathematics, University of Miami, Coral Gables, FL 33124, USA}

%\vspace{-.65cm}

\begin{abstract}
The relative importance of Lagrangian and population dynamics on
spatial pattern formation in the distribution of plankton near the
ocean's surface is investigated. Phytoplankton and zooplankton are
treated as biologically interacting tracers that are passively
advected by a divergence-free two-dimensional nonsteady velocity
field. The time-dependence of this field is assumed to be
quasiperiodic, which extends the analysis presented beyond standard
chaotic advection. Various forms of predator--prey interactions,
including complex interactions, are considered. Numerical
simulations illustrate how Lagrangian dynamics, which is
characterized by a mixed phase space with coexisting regular and
chaotic motion regions, influences plankton patchiness generation
depending on the nature of the biological interactions.

\end{abstract}


\end{frontmatter}

%\vspace{-1cm}

\section{Introduction}

The central role played by the peculiar topology of the phase space
associated with the Lagrangian dynamics of a deterministic divergence-free
two-dimensional oscillatory velocity field in controlling dispersion of
passive tracers on the surface of the ocean was first emphasized by
Pasmanter \citep[][]{Pasmanter-88}. In that work a kinematic model for
advection by a tidal current is used to provide a possible explanation for
the observed \citep[][]{Talbot-74} anomalous (i.e., non-Brownian) diffusion
in diagrams of the average square size of pollutant clouds as a function of
time in shallow seas. Because oscillations of the velocity field in %
\citep[][]{Pasmanter-88} are assumed to be periodic, the Lagrangian motion
is governed by a one-and-a-half-degree-of-freedom Hamiltonian system. The
phase space of this motion, whose coordinates are those of the physical
space for advection, is characterized by the coexistence of regular and
chaotic fluid particle trajectories. More specifically, the phase space
exhibits a mixed structure composed essentially of three types of
topological objects: chaotic motion regions; invariant tori---so-called
\emph{KAM tori} after Kolmogorov, Arnold and Moser---which carry
quasiperiodic motion and separate the chaotic regions; and other invariant
tori---so-called \emph{resonance islands}---which are built up also from
quasiperiodic trajectories and populate the chaotic regions
\citep[cf.,
e.g.,][]{Arnold-89}. The presence of the invariant sets of regular motion
fundamentally influences the dispersion properties of clouds of points
(fluid particles) in phase space \citep[cf., e.g.,][]{Zaslavsky-02} by two
mechanisms. First, both KAM tori and resonance islands serve as impenetrable
barriers for transport. Second, phase-space points can experience stickiness
on the boundary of the resonance islands, and also undergo long excursions
(so-called L\'{e}vy flights) within the chaotic regions. As a consequence,
the dispersion of clouds of phase-space points deviates from a normal (i.e.,
Brownian) diffusive process, in accordance with the observed anomalous
dispersive characteristics of passive tracers.

An important observation is that the above type of Lagrangian dynamics,
which we refer to as \emph{Lagrangian mixed phase-space dynamics}, is not
exclusively restricted to advection by a time-periodic velocity field, which
corresponds to the standard chaotic advection paradigm \citep[][]{Aref-84}.
Rather, velocity fields with complicated time (and also spatial) dependence
have been shown \citep[][]{Beron-etal-04a,Makarov-etal-05,Budyansky-etal-05}
to produce Lagrangian mixed phase-space dynamics as well. By complicated
time dependence we specifically mean quasiperiodic time dependence where the
number of basic frequencies is very large. We remark, however, that the
above observation should not come as a surprise inasmuch as KAM theory %
\citep[cf., e.g.,][]{Arnold-89} has been extended %
\citep[][]{Jorba-Simo-96,Jorba-Villanueva-97} to include near-integrable
nonautonomous Hamiltonians with quasiperiodic time dependence. Such time
dependence arises naturally when a highly structured velocity field is
represented in Fourier space as the superposition of many harmonics, which
may include randomly chosen phases, and whose amplitudes are taken from an
appropriate wavenumber spectrum with some appropriately chosen low- and
high-wavenumber cutoffs. This indicates, for instance, that small-scale
environmental perturbations to a large-scale velocity field can be treated
deterministically without fundamentally changing the Lagrangian mixed
phase-space dynamics resulting from consideration of the large-scale
advection component in isolation. Furthermore, this suggests a
quasi-deterministic representation of mesoscale turbulence, thereby leading
to the expectation that Lagrangian mixed phase-space dynamics may also play
a role in open-ocean environments. The latter expectation finds some support
in the analysis of surface drifter and submerged float trajectories %
\citep[e.g.,][]{Osborne-etal-86a,Osborne-etal-89,Richardson-etal-89,Brown-Smith-90,Nycander-Doos-Coward-02}%
, which have revealed fractal properties.

From the perspective of the distribution of a passive tracer (i.e., any
scalar that does not alter the advection field) the chaotic motion regions
of a mixed phase space correspond to regions of vigorously stirred tracer.
The regular motion sets, particularly the resonance islands, correspond to
patches of poorly stirred tracer. For biologically interacting passive
tracers, such as phytoplankton and zooplankton, the relationship between the
spatial patterns observed in their distributions on the ocean's surface %
\citep[e.g.,][]{Denman-Platt-76} and the features that comprise a mixed
phase space is not obvious. For instance, the effect of fluid particle
clustering in a resonance island may be compensated by the occurrence of
nonequilibrated population dynamics leading to a region of heterogeneously
distributed plankton rather than a patch of homogeneous plankton
concentration. Numerical evidence %
\citep[e.g.,][]{Abraham-98,Bees-etal-98,Neufeld-etal-02,Martin-03,Reigada-etal-03,Vilar-etal-03}
nonetheless suggests that advection by ocean currents can indeed play an
important role in controlling the generation of spatial patterns in plankton
distributions. However, the numerical evidence presented have typically
relied on either the assumption of equilibrated population dynamics or the
consideration of a specific biological interaction scenario. So far as we
are aware, no work has been reported on plankton patchiness generation that
focuses on the competing roles played by Lagrangian mixed phase-space
dynamics and the predator--prey dynamics controlling the plankton density
intrinsic evolution.

The main goal of the present contribution is to carry out such a study. To
keep our analysis simple, we will consider a kinematic model of advection by
a tidal current that is only somewhat more complicated than that used in %
\citep[][]{Pasmanter-88}. More precisely, we will take into account the
spring-to-neap cycle associated with the superposition of the semilunar
(semidiurnal lunar) and semisolar (semidiurnal solar) tides, which leads to
a streamfunction (Hamiltonian) that depends quasiperiodically on time.
Consideration of this time dependence will extend the scope of our work
beyond the realm of standard chaotic advection, which differs in some
respects from advection by a time-quasiperiodic velocity field. A secondary
goal of the present work is discuss these differences, which do not depend
on the number $n$ of basic frequencies considered, thereby allowing us to
choose $n=2$ with no loss of generality. Unlike prior studies of plankton
patchiness generation, we will consider a variety of biological interaction
types, including complex interactions, which will be assumed to be governed
by a simple set of predator--prey equations. Specifically, in addition to
autonomous predator--prey interactions we will also consider nonautonomous
predator--prey dynamics by allowing for a diurnal cycle in the food
availability to prey, which will lead to a very rich set of interaction
scenarios, including chaotic predator--prey interactions. Remarkably, even
though the biological relevance of the latter interactions is supported both
theoretically and experimentally
\citep[][and references
therein]{Scheffer-etal-03}, they have been excluded in prior studies of the
production of plankton patchiness. The rationale for their exclusion may be
that their presence does not lead to the development of spatial patterns in
the plankton distributions. However, anticipating some of our results, in
this paper we will show that Lagrangian mixed phase-space dynamics can lead
to plankton patchiness development even under chaotic predator--prey
interactions. It will be shown that in general the influence of Lagrangian
mixed phase-space dynamics plays on plankton patchiness generation is less
constrained by the type of biological interactions during early stages of
the evolution than during later stages.

The remainder of the paper has the following organization. In Section 2 we
introduce the time-quasiperiodic kinematic velocity field model that is used
throughout the paper. In the same section we discuss the effects of the
time-quasiperiodicity of the velocity field on Lagrangian dynamics. This is
done by suitably constructing a Poincar\'{e} section. Section 3 discusses
the dynamics of tracers passively advected by the velocity field. Section 4
is devoted to study the dynamics of phytoplankton and zooplankton by
treating them as biologically interacting tracers that are passively
advected by the velocity field defined earlier. The different types of
biological interactions considered, both autonomous and nonautonomous, are
discussed prior to discussing the effects of advection. Section 5 offers a
discussion on several topics including the effects of random fluctuations of
the velocity field and the relevance of the present work to the study of
plankton advection by more complicated ocean currents. A summary and the
conclusions of the paper are finally presented in Section 6.

\section{Fluid Particle Dynamics}

We consider a divergence-free velocity field $\mathbf{u}=(\partial _{y}\psi
,-\partial _{x}\psi )$ defined at each time $t$ on the Euclidean plane with
Cartesian coordinates $(x,y)$ by a streamfunction $\psi $ of the form %
\citep[][]{Pasmanter-88}
\begin{equation}
\psi (x,y,t):=u(t)y-v(t)x+U(t)l^{-1}\cos ly-V(t)k^{-1}\cos kx,  \label{psi}
\end{equation}%
where%
\begin{equation}
u(t)=u_{0}+\sum_{1}^{n}u_{i}\cos \sigma _{i}t
\end{equation}%
and similarly for $v(t),$ $U(t),$ and $V(t)$. The velocity field $\mathbf{u}$
includes the essential features of a tidal (i.e., oscillatory) velocity
field in a shallow sea wherein ebb and flood surpluses organize in cell-like
circulation structures. A dynamical explanation of these \textquotedblleft
tidal residual eddies\textquotedblright\ can be found in %
\citep[][]{Zimmerman-78,Zimmerman-81}. In \citep[][]{Pasmanter-88} the
kinematic model (\ref{psi}) is written for arbitrary $n$ with the
frequencies $\{\sigma _{i}\}$ chosen to be commensurate, which leads to
periodic time dependence. Analysis in \citep[][]{Pasmanter-88} is restricted
to the $n=1$ case, which corresponds to the standard chaotic advection
paradigm. A model similar to (\ref{psi}) with $n=1$ is also used in %
\citep[][]{Ridderinkhof-Zimmerman-92}. In this paper we consider a
time-quasiperiodic dependence with $n=2$ incommensurate frequencies. This
choice presents all the mathematical difficulties associated with a large
number of incommensurate frequencies, and furthermore extends our analysis
beyond standard chaotic advection.

More precisely, we choose $\sigma _{1}=2\pi /12.4206$ rad h$^{-1}$ and $%
\sigma _{2}=2\pi /12$ rad h$^{-1}$, which presumes a primarily semidiurnal
tidal regime that includes the spring-to-neap tidal cycle associated with
the superposition of the semilunar tidal component and a smaller semisolar
tidal component. For this frequency choice $\mathbf{u}$ is actually
time-periodic with a period of exactly $1495224$ h ($\approx 170$ y);
however, this period is too long and $\mathbf{u}$ is more appropriately
interpreted as being time-quasiperiodic. In %
\citep[][]{Ridderinkhof-Zimmerman-92} it is speculated about the possible
similarities of the effects of this time-quasiperiodicity on fluid particle
motion compared to the effects of time-periodicity. In this paper we show
that there indeed are qualitative similarities but also quantitative
differences.

The trajectories $(x(t),y(t))$ of fluid particles obey%
\begin{equation}
\dot{x}=\partial _{y}\psi ,\quad \dot{y}=-\partial _{x}\psi .  \label{xy}
\end{equation}%
Upon identifying $x$--$y$ with the canonically conjugate
coordinate--momentum pair, system (\ref{xy}) is seen to constitute a
one-degree-of-freedom Hamiltonian system that is nonautonomous because the
Hamiltonian, $\psi $, depends explicitly on $t$. More specifically, such
dependence is assumed of the form%
\begin{equation}
\psi (x,y,t)=\Psi (x,y,\sigma _{1}t,\sigma _{2}t),  \label{Psi}
\end{equation}%
where $\Psi :\mathbb{R}^{2}\times \mathbb{T}^{2}\rightarrow \mathbb{R}^{2}$ (%
$\mathbb{T}=\mathbb{R}/2\pi \mathbb{Z}$ is the standard one-torus). The
phase space of (\ref{xy}) is constituted by the real two-space $\mathbb{R}%
^{2}$ endowed with the standard symplectic form $\mathrm{d}x\wedge \mathrm{d}%
y,$ which is preserved by the time-dependent flow of $\mathbf{u}$. Namely,
the family of $(t,s)$-advance mappings of $\mathbb{R}^{2}$ to itself%
\begin{equation}
\phi _{t,s}:(x(t),y(t))\mapsto (x(s),y(s))  \label{phi}
\end{equation}%
satisfying $\phi _{t,s}\circ \phi _{s,s+r}=\phi _{t,s+r}$ and $\phi _{t,t}=%
\limfunc{id}$. Invariance of $\mathrm{d}x\wedge \mathrm{d}y$ under the flow
of $\mathbf{u}$ is equivalent to preservation of area in phase space
(Liouville's theorem).

\begin{figure}[t]
\centering
\includegraphics[width=12cm]{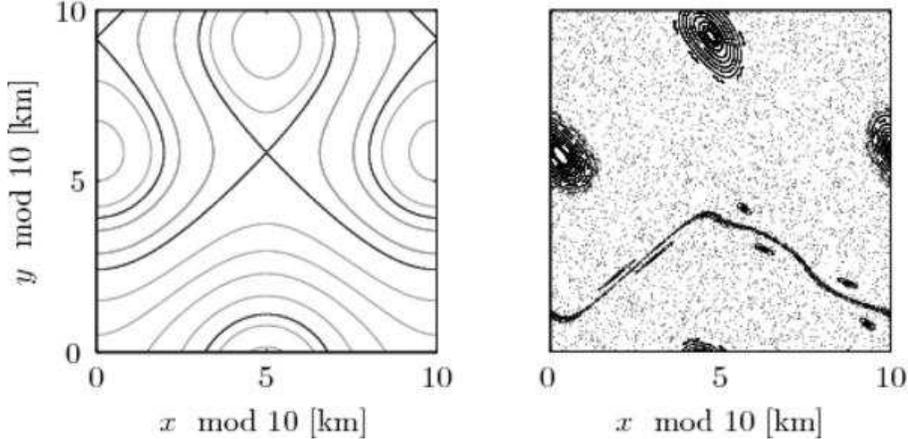}
\caption{Phase portraits of (\protect\ref{xy}) with $\protect\psi $ defined
by (\protect\ref{psi}) excluding time-dependent components (left) and
including two time-dependent components (right) with frequencies, which, for
any practical purpose, can be interpreted as being incommensurate. Parameter
choices are: $\protect\sigma _{1}=2\protect\pi /12.4206$ rad h$^{-1}$, $%
\protect\sigma _{2}=2\protect\pi /12$ rad h$^{-1},$ $k=l=2\protect\pi /10$ km%
$^{-1}$, $u_{0}=-0.5$ km d$^{-1},$ $v_{0}=0$ km d$^{-1},$ $U_{0}=V_{0}=1$ km
d$^{-1}$, $u_{1}=v_{1}=u_{2}=v_{2}=-0.1$ km d$^{-1},$ and $%
U_{1}=V_{1}=U_{2}=V_{2}=0.1$ km d$^{-1}$. The plot in the right panel
corresponds to a second-order Poincar\'{e} section of (\protect\ref{xy})
defined by (\protect\ref{Sigma}) with $\protect\delta =10^{-4}.$}
\label{Poincare}
\end{figure}

The time-independent part of $\psi $ is integrable; typical fluid particle
trajectories are depicted in the left panel of Fig. \ref{Poincare} on $%
\mathbb{R}/2\pi k^{-1}\mathbb{Z}\times \mathbb{R}/2\pi l^{-1}\mathbb{Z}$. In
general, due to explicit time dependence, $\psi $ is not integrable and
trajectories may be chaotic, i.e., exhibit sensitivity to initial
conditions. A two-dimensional portrait of the dynamics can be sketched by
constructing a second-order Poincar\'{e} section\
\citep[cf., e.g.,][Chapter
2.3]{Parker-Chua-89}. This is accomplished by taking the first return map
under the flow of (\ref{xy}) to the set%
\begin{equation}
\Sigma :=\mathbb{R}^{2}\times \left\{ \varphi _{1}=0,\text{ }\varphi _{2}\in %
\left[ 0,\delta \right] \right\}   \label{Sigma}
\end{equation}%
where $\varphi _{i}:=\sigma _{i}t$ $\func{mod}2\pi $ and  $\delta >0$ is
small, and then plotting the resulting points using their $\mathbb{R}^{2}$
coordinates.\footnote{%
Because $\delta $ can never be chosen to be equal to zero, the set $\Sigma $
actually lies on a three-dimensional space. Consequently, a second-order
Poincar\'{e} section is not a plane in a strict sense; rather, it is a slice
of nonzero, albeit thin, thickness.}

The right panel of Fig. \ref{Poincare} shows a second-order Poincar\'{e}
section of (\ref{xy}) assuming $\delta =10^{-3}$. To ensure strict
Hamiltonian behavior, the flow of (\ref{xy}) was numerically computed using
a fully-implicit sixth-order Gauss--Legendre method
\citep[cf.,
e.g.,][]{McLachlan-Atela-92}, which is symplectic (i.e., area preserving).
Despite the blurring associated with finiteness of $\delta $, the right
panel of Fig. \ref{Poincare} reveals quite clearly a mixed phase space with
coexisting regular and chaotic motion regions. Such a mixed phase space is
most commonly associated with Poincar\'{e} sections of (\ref{xy}) with
time-periodic $\psi $ constructed by strobing trajectories at the frequency
of $\psi $. Legitimacy of the mixed phase-space topology identified here
with the aid of the second-order Poincar\'{e} section technique is insured
by the rigorous KAM-theory-related results of Jorba and Sim\'{o} %
\citep[][]{Jorba-Simo-96}.

To grasp the meaning of the mixed phase-space dynamics associated with (\ref%
{xy}) for time-quasiperiodic $\psi $, it is most convenient to consider $%
\psi $ split into background, $\psi _{0}(I)$, and perturbation, $\psi
_{1}(I,\vartheta ,t)+\psi _{2}(I,\vartheta ,t)$ where $\psi _{i}(I,\vartheta
,t)=\Psi _{i}(I,\vartheta ,\sigma _{i}t).$ Here $I$--$\vartheta $ are some
locally defined action--symplectic-angle variables. When $\psi _{1}+\psi
_{2}=0,$ motion lies on one-tori $\{I=\func{const}\}$ with frequency $\omega
(I):=\psi _{0}^{\prime }(I)$ (lighter curves in the left panel of Fig. \ref%
{Poincare}). When $|\psi _{1}+\psi _{2}|\neq 0$ is sufficiently small, the
KAM theorem proved in \citep[][]{Jorba-Simo-96} guarantees that the
background tori for which the frequencies $\{\omega (I),\sigma _{1},\sigma
_{2}\}$ are incommensurate in a Diophantine sense survive the perturbation,
albeit slightly distorted and with the frequencies $\sigma _{1}$ and $\sigma
_{2}$ added to the unperturbed frequency $\omega (I)$. The latter result,
which applies to almost all background tori, holds under the assumptions
that $\psi $ is sufficiently smooth, $\sigma _{1}$ and $\sigma _{2}$ satisfy
a Diophantine inequality, and $\psi _{0}$ is nondegenerate in the sense $%
\omega ^{\prime }(I)\neq 0$.\footnote{%
However, there are conditions \citep[e.g.,][]{Sevryuk-95a} on higher-order
derivatives of $\omega (I)$ that probably \citep[][]{Rypina-etal-06} allow
to make density estimates of tori even in the neighborhood of a point where $%
\omega ^{\prime }(I)=0$.} The main effect of the perturbation is to impart
on a background torus a quasiperiodic vibration with frequencies $\sigma _{1}
$ and $\sigma _{2}$, which has been referred in \citep[][]{Jorba-Simo-96} to
as a \textquotedblleft quasiperiodic dancing.\textquotedblright\ This
contrasts with the widely studied time-periodic perturbation case for which
the vibration is periodic.

The open and concentric curves in the right panel of Fig. \ref{Poincare} are
examples of KAM tori\footnote{%
More precisely, cantori, due to insufficient incommensurability of $\sigma
_{1}$ and $\sigma _{2}$. Cantori are gappy tori (more rigorously, Cantor
sets) on which the dynamics is similar to that on KAM tori, except that they
can be occasionally traversed by trajectories through the gaps %
\citep[][]{Aubry-78,Percival-79}.} of the above quasiperiodically vibrating
type. The closed curves flanking the KAM tori correspond to tori into which
background tori have broken up due to resonance, e.g., $\omega
(I_{0})/\sigma _{1}=p/q\in \mathbb{Q}$. These tori (so-called resonance
islands) enclose elliptic points separated by an equal number of saddle
points, which, unlike the well-known time-periodic perturbation case,
correspond to quasiperiodic trajectories, e.g., with frequencies $\{q\omega
(I_{0})=p\sigma _{1},\sigma _{2}\}$. Associated with the saddle points are
stable and unstable manifolds that transversely intersect infinitely many
times producing a narrow band of chaotic motion (so-called chaotic layer) as
do perturbed heteroclinic trajectories \citep[e.g.][]{Beigie-etal-91}. The
fuzzy two-dimensional regions filled up with dots in the right panel of Fig. %
\ref{Poincare} correspond to chaotic motion regions. These regions originate
from the merging of the chaotic layers associated with the breakup of two
homoclinic background trajectories (darker curves in the left panel of Fig. %
\ref{Poincare}) and neighboring chaotic layers associated with the breakup
of resonant background tori that are not separated by KAM tori.

The above results straightforwardly extend \citep[][]{Jorba-Simo-96} to the
case of an arbitrary number $n$ of perturbation frequencies. When $n$ is
large, however, other phase-space visualization techniques must be
considered. Examples are the estimation of Lyapunov exponents and a variant
of the Poincar\'{e} section technique designed in %
\citep[][]{Makarov-Uleysky-06} to detect domains of finite-time stability in
the phase space of randomly perturbed systems. In \citep[][]{Beron-etal-04a}
two-dimensional charts of finite-time estimates of Lyapunov exponents were
constructed for the Lagrangian motion produced by a gyre-like ocean
circulation composed of a steady component and a unsteady perturbation with
a broad band of frequencies. These charts revealed regions of small absolute
numerical values, suggesting KAM tori and resonance islands, and regions of
larger absolute numerical values, identifiable with chaotic motion regions.
Similar results were found \citep[][]{Makarov-etal-05,Budyansky-etal-05}
using the Poincar\'{e} section technique of \citep[][]{Makarov-Uleysky-06}
applied to the Lagrangian motion produced by an oceanic jet interacting with
an eddy in a noisy environment. This numerical evidence suggests that the
results presented below also apply to the large $n$ case, which allows to
represent more complicated (e.g., turbulent) advection fields (cf. Section
4).

\section{Passive Tracer Dynamics}

The concentration of a passive tracer $\theta (x,y,t)$ evolves according to
the Liouville (advection) equation%
\begin{equation}
\partial _{t}\theta +[\theta ,\psi ]=0,  \label{L}
\end{equation}%
where $[f,g]:=\partial _{x}f\partial _{y}g-\partial _{y}f\partial _{x}g$
denotes the symplectic bracket with respect to $\mathrm{d}x\wedge \mathrm{d}%
y $. Equation (\ref{L}) is equivalent to $\dot{\theta}=0$ along $(x(t),y(t))$
satisfying (\ref{xy}), i.e., $\dot{x}=[x,\psi ]$ and $\dot{y}=[y,\psi ]$.
Consequently, the solution of (\ref{L}) for certain initial distribution $%
\theta (x,y,0)$ may be expressed as%
\begin{equation}
\theta (x,y,t)=\theta (\phi _{0,-t}(x,y\mathbf{)},0).
\end{equation}%
In other words, the distribution of $\theta $ at any instant is an
area-preserving rearrangement of its initial distribution.

Existence and uniqueness of solutions of (\ref{xy}) guarantee that
trajectories with different initial conditions can never intersect. KAM tori
and resonance islands are invariant objects in the sense that they are built
up from trajectories; hence, they can never be traversed by trajectories
with different initial conditions. In fluid mechanics jargon, these objects
are said to be material meaning that they are composed by the same fluid
particles. The chaotic stirring of a passive tracer is thus constrained by
KAM tori, which are barriers for tracer transport, and resonance islands,
which constitute actual tracer patches. Both types of mixed space-space
structures lead to spatial pattern formation in the continuous-time
distribution of a passive tracer.
\begin{figure}[t]
\centering
\includegraphics[width=13.25cm]{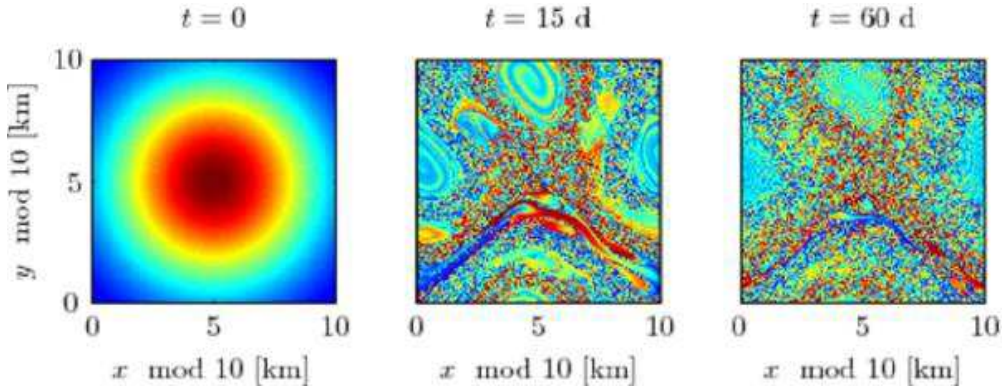}
\caption{Snapshots of the evolution of a tracer that is passively advected
by the time-quasiperiodic tidal velocity field defined by streamfunction (%
\protect\ref{psi}) with the same parameters used to construct the Poincar%
\'{e} section in the right panel of Fig. \protect\ref{Poincare}. The color
scale indicates tracer concentration or simply a fluid particle labeling.}
\label{Tracer}
\end{figure}

Fig. \ref{Tracer} illustrates the above for an arbitrary tracer that is
passively advected by the quasiperiodic tidal current defined by
streamfunction (\ref{psi}) with parameters as in Fig. \ref{Poincare}. To
construct Fig. \ref{Tracer} we advected $250000$ fluid particles distributed
initially over a regular grid defined on a doubly-periodic $10$ km $\times $
$10$ km domain of $\mathbb{R}^{2}$. The spatial distribution of the tracer
at any stage of the evolution is obtained from Delaunay triangulation of
fluid particle positions by triangle-based cubic interpolation onto the
regular grid. The distribution of the tracer at $t=0$ is assumed to be a
Gaussian. At $t=15$ d the tracer distribution reveals vigorously stirred
regions along with poorly stirred regions in the form of spiral-like
vortices and filaments. These features resemble quite closely those
embodying the mixed phase space shown in the right panel of Fig. \ref%
{Poincare}. The mixed phase-space features are much more clearly visible in
the tracer distribution at the later instant $t=60$ d, which can be
considered long enough for (constrained) chaotic stirring to develop
substantially. Note, however, that $60$ d is much shorter than the
integration time required to construct the second-order Poincar\'{e} section
in the right panel of Fig. \ref{Poincare}.

An important observation is that the tracer distribution features can never
match those of the mixed phase space because, as explained above, the mixed
phase space features \textquotedblleft dance\textquotedblright\ with the
\textquotedblleft rhythm\textquotedblright\ of the quasiperiodic tide. This
contrasts with the standard chaotic advection paradigm (time-periodic $\psi $%
) for which the motion of the mixed phase space structures\ is periodic. In
that case matching can be achieved by displaying the tracer distribution at
some (sufficiently large) multiple of the time-period of $\psi $.

\section{Biologically Interacting Passive Tracer Dynamics}

This section is devoted to show that Lagrangian mixed phase-space dynamics
can play a distinguished role in setting the distribution of biologically
interacting tracers depending on the nature of the interaction. We will
assume that the biologically interacting tracers obey a coupled set of
Liouville equations of the form%
\begin{equation}
\partial _{t}\theta +[\theta ,\psi ]=F_{\theta }(\theta ,t),  \label{AT}
\end{equation}%
where $\theta $ is understood as a vector of biologically interacting
tracers and $F_{\theta }(\theta ,t)$ represents the interaction. The
solution of (\ref{AT}) may be expressed as%
\begin{equation}
\theta (x,y,t)=\theta _{\mathrm{rest}}(\phi _{0,-t}(x,y),t),
\end{equation}%
where $\theta _{\mathrm{rest}}(x,y,t)$ is the solution of (\ref{AT}) with
initial distribution $\theta _{\mathrm{rest}}(x,y,0)=\theta (x,y,0)$
assuming that the fluid is at rest. In other words, the distribution of $%
\theta $ at time $t$ is given by an area-preserving rearrangement of the
distribution it would have attained at that same instant if the fluid were
at rest.

Our particular focus will be on the plankton distribution problem, which we
will approach by considering the interaction between phytoplankton, $P$, and
zooplankton, $Z$, in a two-level food chain. We will consider nonautonomous
Rosenzweig--MacArthur \citep[][]{Rosenzweig-MacArthur-63} predator--prey
interactions governed by%
%TCIMACRO{\TeXButton{inicio letras}{\begin{mathletters}\label{F}}}%
%BeginExpansion
\begin{mathletters}\label{F}%
%EndExpansion
\begin{gather}
F_{P}(P,Z,t)=r(t)\left( 1-\frac{P}{K(t)}\right) P-\frac{cPZ}{P_{\mathrm{s}}+P%
},  \label{Fp} \\
\hspace{-2.9cm}F_{Z}(P,Z,t)=\frac{ecPZ}{P_{\mathrm{s}}+P}-dZ,  \label{Fz}
\end{gather}%
%TCIMACRO{\TeXButton{fin mathletters}{\end{mathletters}}}%
%BeginExpansion
\end{mathletters}%
%EndExpansion
where%
\begin{equation}
r(t)=r_{0}(1-\epsilon \cos \Omega t),\quad K(t)=K_{0}(1-\epsilon \cos \Omega
t).  \label{rK}
\end{equation}%
In this model the prey density, $P$, grows logistically with growth rate $%
r(t)>0$ and carrying capacity $K(t)>0$ in the absence of predation, $Z=0$.
The predator consumes the prey according to a Holling's type-II functional
response, which presumes a maximal consumption rate $c>0$ and
half-saturation density $P_{\mathrm{s}}>0$. The model further assumes that
the predator's per capita death rate, $d>0$, is independent of density, and
that its numerical response is proportional to its functional response with
efficiency $e>0.$ We will set $2\pi /\Omega =1$ d so that (\ref{rK}) crudely
accounts for the diurnal cycle of light availability.

Other, more sophisticated and perhaps more realistic, two- or higher-level
food chain models and multispecies systems have been proposed, which could
of course be considered. However, model (\ref{F}), while very simple,
describes a very rich variety of biological interaction types, which is
enough for the purposes of the present work.

\subsection{Biological Interactions}

The predator--prey interactions governed by (\ref{F}), i.e., $\dot{P}%
=F_{P}(P,Z,t)$ and $\dot{Z}=F_{Z}(P,Z,t)$ along $(x(t),y(t))$, constitute a
dissipative dynamical systems provided $P_{\mathrm{e}}/K\neq 0$ and $P_{%
\mathrm{e}}/P_{\mathrm{s}}\neq 0,$ where
\begin{equation}
P_{\mathrm{e}}:=\frac{dP_{\mathrm{s}}}{ec-d},  \label{Pe}
\end{equation}%
which is biologically most sensible. Consequently, the topology of phase
space for predator--prey interactions will be fundamentally different than
the topology of phase space for fluid particle motion governed by (\ref{xy})
and discussed in Section 2. Predator--prey interactions are described in the
next paragraph; for a more detailed description the reader is referred to
Refs. \citep[][]{Rinaldi-etal-93,King-Schaffer-99}.

In the autonomous case ($\epsilon =0$), predator--prey interactions obeying (%
\ref{F}) possess a unique positive equilibrium $(P_{\mathrm{e}},Z_{\mathrm{e}%
}),$ where%
\begin{equation}
Z_{\mathrm{e}}:=\frac{r_{0}}{c}\left( 1-\frac{P_{\mathrm{e}}}{K_{0}}\right)
\left( P_{\mathrm{s}}+P_{\mathrm{e}}\right) ,  \label{Ze}
\end{equation}%
provided $0\leq P_{\mathrm{e}}/K_{0}<1$ and $P_{\mathrm{e}}/P_{\mathrm{s}%
}\geq 0$. This equilibrium is subject to a change of stability type through
a Hopf bifurcation. For $P_{\mathrm{e}}/K_{0}>1/(2+P_{\mathrm{s}}P_{\mathrm{e%
}})$ predator and prey densities undergo damped oscillations toward $(P_{%
\mathrm{e}},Z_{\mathrm{e}})$, while for $P_{\mathrm{e}}/K_{0}<1/(2+P_{%
\mathrm{s}}/P_{\mathrm{e}})$ predator and prey densities experience
time-asymptotic, self-sustained oscillations about $(P_{\mathrm{e}},Z_{%
\mathrm{e}})$.

\begin{figure}[t]
\centerline{\includegraphics[width=12cm]{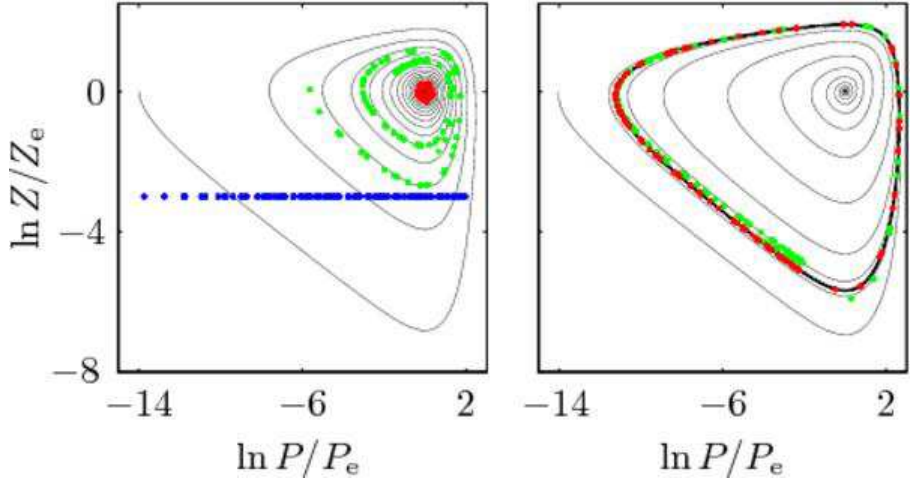}}
\caption{Phase portraits for phytoplankton, $P$, and zooplankton, $Z$,
interacting according to autonomous predator--prey dynamics governed by (%
\protect\ref{F}), i.e., with $\protect\epsilon =0$ in (\protect\ref{rK}).
The thin-solid curve in the right panel depicts a damped oscillation toward $%
(P_{\mathrm{e}},Z_{\mathrm{e}})$, defined in Eqs. (\protect\ref{Pe})--(%
\protect\ref{Ze}), for $\ln P_{\mathrm{e}}/K_{0}=-3$ and $\ln P_{\mathrm{e}%
}/P_{\mathrm{s}}=-4$. The thick-solid closed curve in the left panel
illustrates a self-sustained oscillation about the equilibrium $(P_{\mathrm{e%
}},Z_{\mathrm{e}})$ for $\ln P_{\mathrm{e}}/K_{0}=-2.8$ and $\ln P_{\mathrm{e%
}}/P_{\mathrm{s}}=-1.6$; orbits for two different initial conditions are
shown in this panel as thin-solid curves, which collapse into the limit
cycle represented by the thick-solid curve. In both panels $r_{0}/\Omega
=0.4475$ and $d/\Omega =0.2011.$ The blue dots in the left panel correspond
to a discrete set of $P$--$Z$ values in the range spanned by the
distributions shown in Figs. \protect\ref{P_t}--\protect\ref{Z_t} at $t=0$.
The green (resp., red) dots in the left and right panels correspond to a
discrete set of $P$--$Z$ values in the range spanned by the distributions
shown, respectively, in the top and mid-top row panels of Figs. \protect\ref%
{P_t}--\protect\ref{Z_t} at $t=15$ d (resp., $t=60$ d).}
\label{PZ_1}
\end{figure}

\begin{figure}[t]
\centerline{\includegraphics[width=12cm]{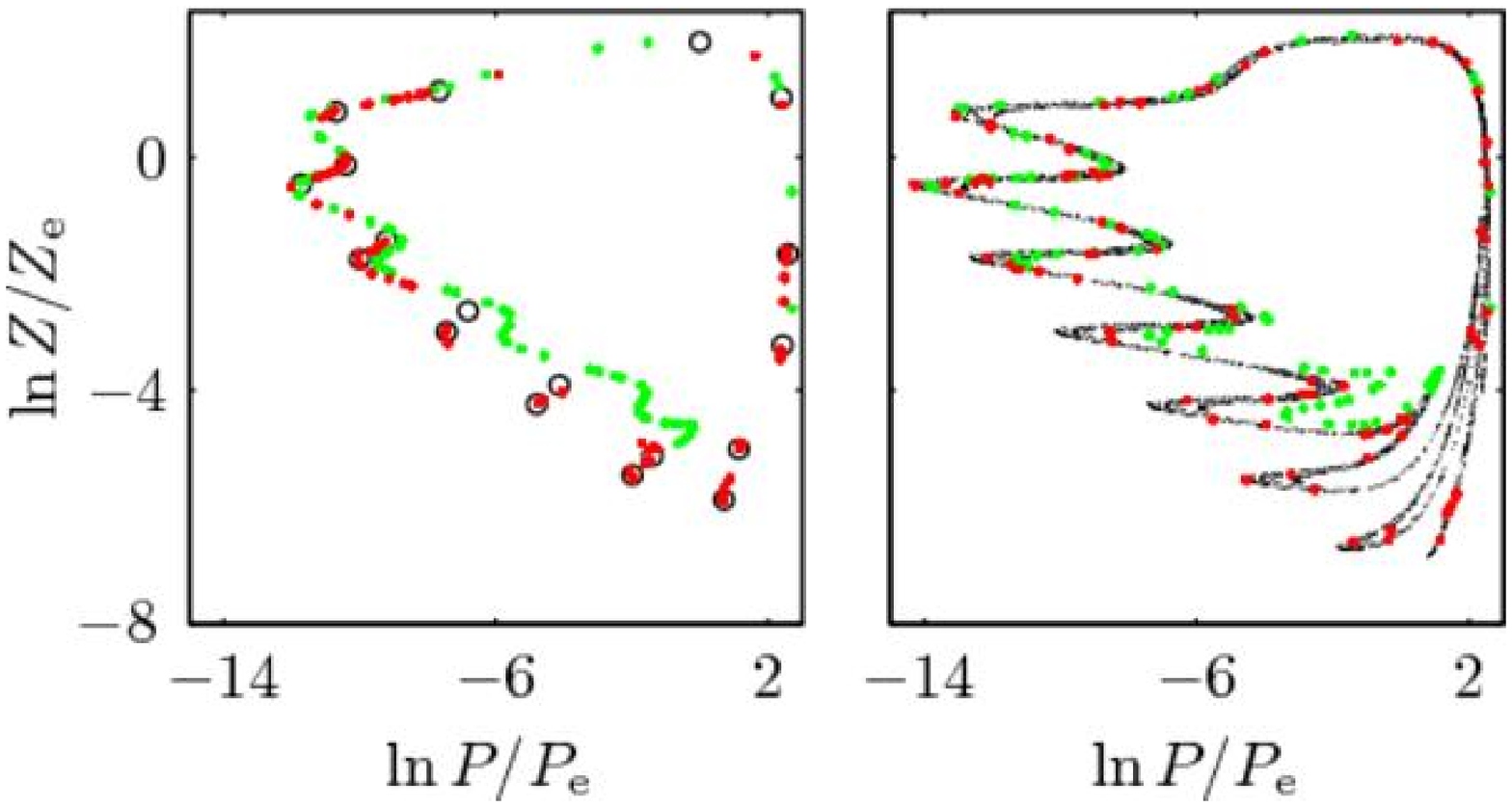}}
\caption{As in Fig. \protect\ref{PZ_1}, but for nonautonomous predator--prey
dynamics governed by (\protect\ref{F}), i.e., with $\protect\epsilon \neq 0$
in (\protect\ref{F}b). The left and right panels correspond to Poincar\'{e}
sections showing, respectively, a periodic attractor ($1:18$ subharmonic)
attained with $\protect\epsilon =0.2$ and a strange attractor attained with $%
\protect\epsilon =0.6;$ in both panels $\ln P_{\mathrm{e}}/K_{0}=-2.8$, $\ln
P_{\mathrm{e}}/P_{\mathrm{s}}=-1.6$, $r_{0}/\Omega =0.4475$, and $d/\Omega
=0.2011.$ The green (resp., red) dots in the left and right panels
correspond to a discrete set of $P$--$Z$ values in the range spanned by the
distributions shown, respectively, in the mid-bottom and bottom row panels
of Figs. \protect\ref{P_t}--\protect\ref{Z_t} at $t=15$ d (resp., $t=60$ d).}
\label{PZ_2}
\end{figure}

In the nonautonomous case ($\epsilon >0$), the equilibrium $(P_{\mathrm{e}%
},Z_{\mathrm{e}})$ is replaced by a small-amplitude periodic orbit that
loses stability through discrete Hopf or Neimark--Sacker bifurcations.
Chaotic predator--prey interactions, which are characterized by the presence
of a strange attractor, eventually give rise via either torus destruction
(quasiperiodic route to chaos) or a cascade of period doublings (subharmonic
route to chaos). The first kind of strange attractors results for small $%
\epsilon $ when the autonomous predator--prey interactions lie on a limit
cycle. Strange attractors of the second class develop for larger $\epsilon $%
, even when the autonomous predator--prey system does not cycle.

Fig. \ref{PZ_1} depicts phase portraits for autonomous interactions that
decay toward a stable spiral equilibrium (left panel) and asymptotically
cycle (right panel). Fig. \ref{PZ_2} shows phase portraits for nonautonomous
interactions taking place on a periodic $1:18$ subharmonic attractor (left
panel) and lying on a strange attractor attained via quasiperiodic route
(right panel). The latter are shown on Poincar\'{e} sections of the
predator--prey system, which are computed by strobing orbits at the
environmental variability frequency $\Omega /2\pi $. To construct Figs. \ref%
{PZ_1}--\ref{PZ_2} the predator--prey equations were numerically integrated
using the Dormand--Prince 5(4) algorithm. The period of the limit cycle
oscillation in the right panel of Fig. \ref{PZ_1} is of about $9$ d.
Estimated Lyapunov exponents scaled by $\Omega $ for the strange attractor
in the right panel of Fig. \ref{PZ_2} are $\{+0.0239,0,-0.0539\}$. This
gives a predictability horizon (inverse of the positive Lyapunov exponent)
not exceeding 10 d for the chaotic interactions considered here.

Other, intermediate, forms of interaction (not shown but discussed below)
are nonautonomous interactions lying on a torus or quasiperiodic attractor,
and interactions possessing multiple periodic or quasiperiodic attractors.
The former are excited for sufficiently small $\epsilon $ when the
autonomous interactions take place on a limit cycle; the latter precede
chaotic interactions.

\subsection{Biological Interactions and Passive Advection}

\begin{figure}[t]
\centering
\subfigure{\includegraphics[width=14cm]{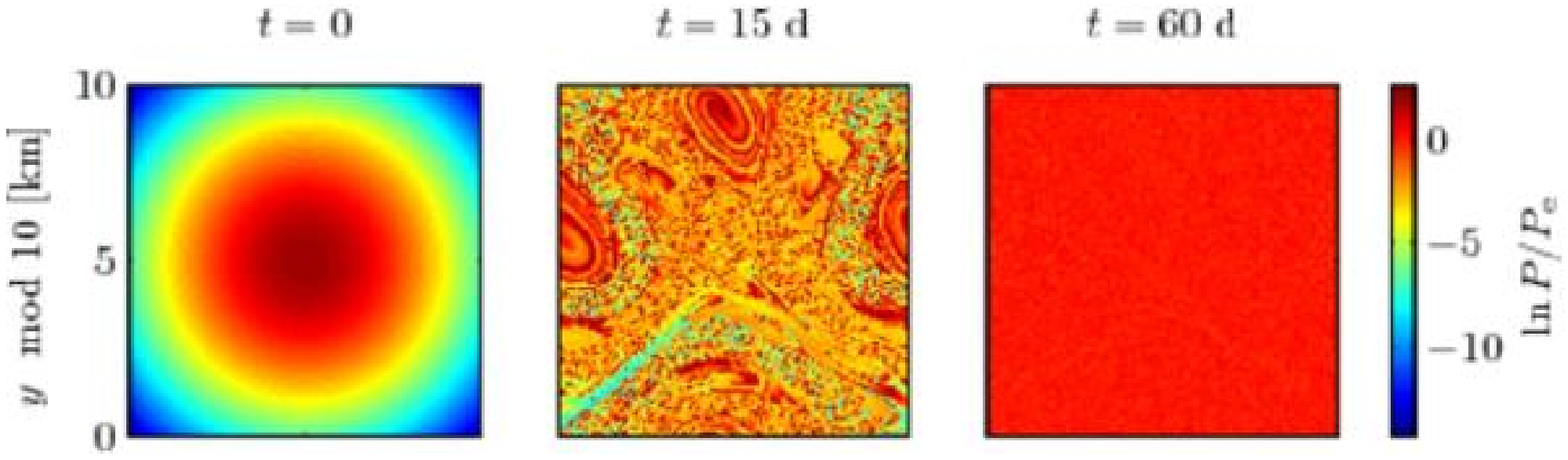}}\newline
\subfigure{\includegraphics[width=14cm]{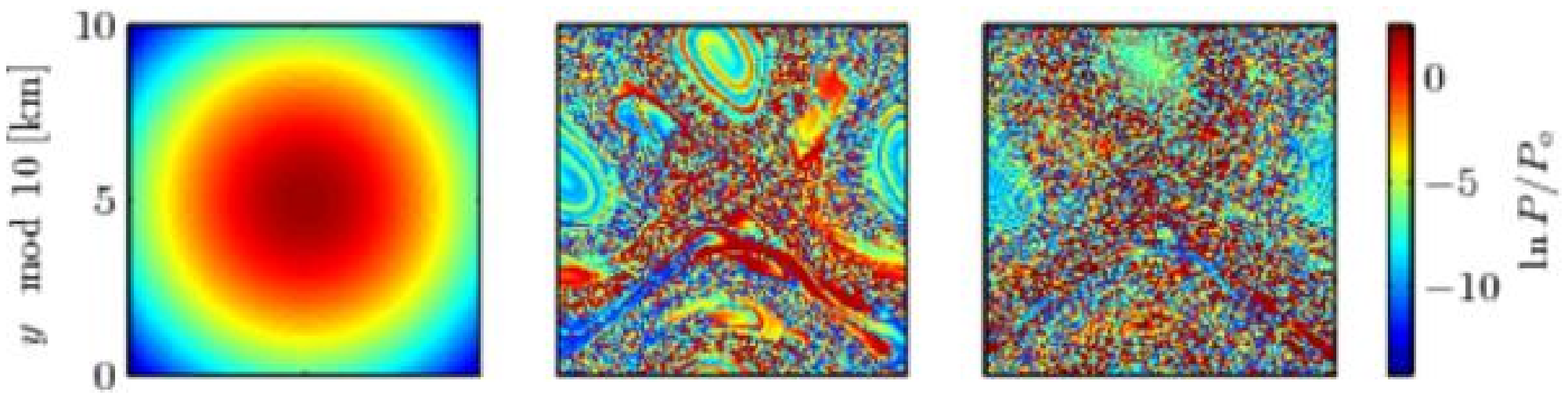}}\newline
\subfigure{\includegraphics[width=14cm]{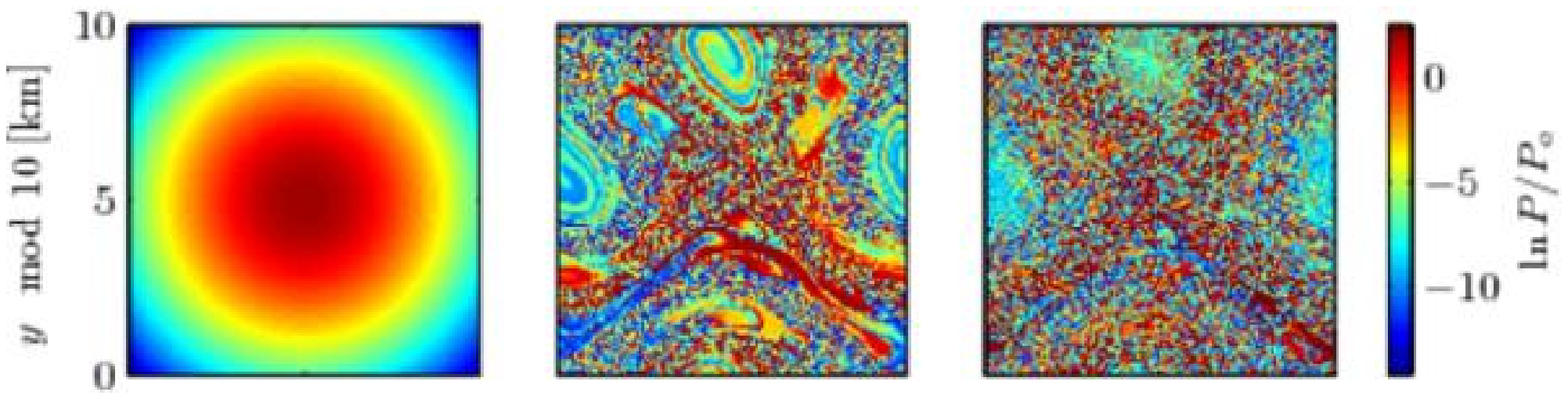}}\newline
\subfigure{\includegraphics[width=14cm]{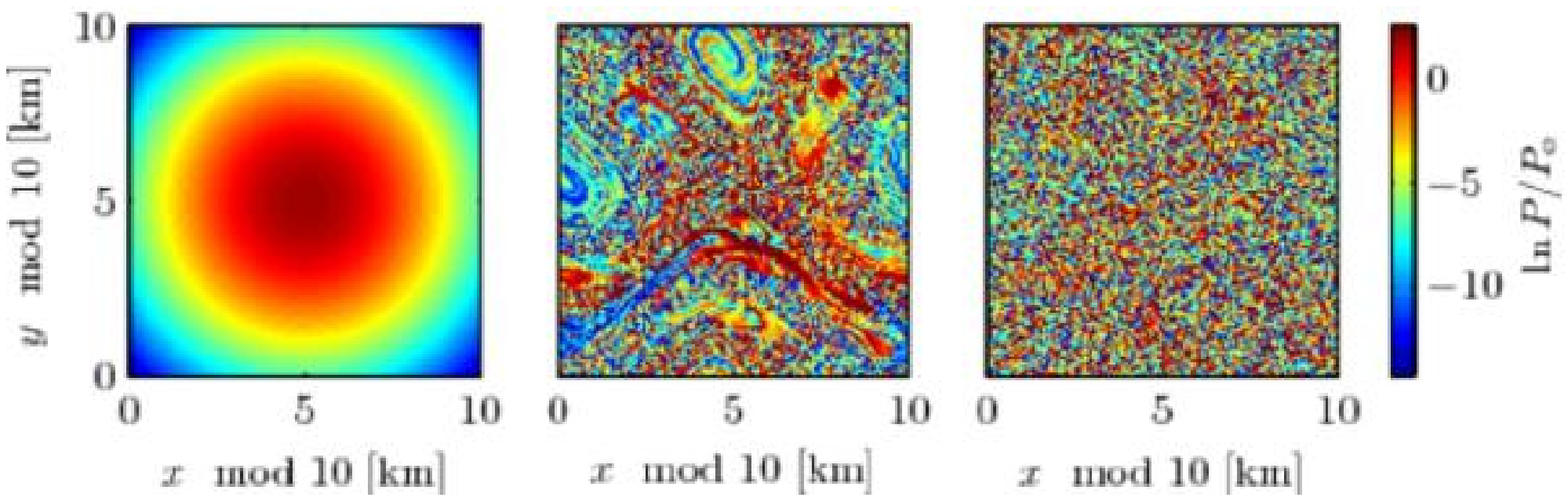}}
\caption{Snapshots of the evolution of phytoplankton, which, while
interacting with zooplankton through predator--prey dynamics governed by (%
\protect\ref{F}), is advected by the time-quasiperiodic tidal velocity field
defined by streamfunction (\protect\ref{psi}) with parameters as in the
right panel of Fig. \protect\ref{Poincare}. Predator--prey interaction
parameters in the top and mid-top panels are, respectively, as in the left
and right panels of Fig. \protect\ref{PZ_1}. Parameters in the mid-bottom
and bottom row panels are, respectively, as in left and right panels of Fig.
\protect\ref{PZ_2}.}
\label{P_t}
\end{figure}

\begin{figure}[t]
\centering\subfigure{\includegraphics[width=14cm]{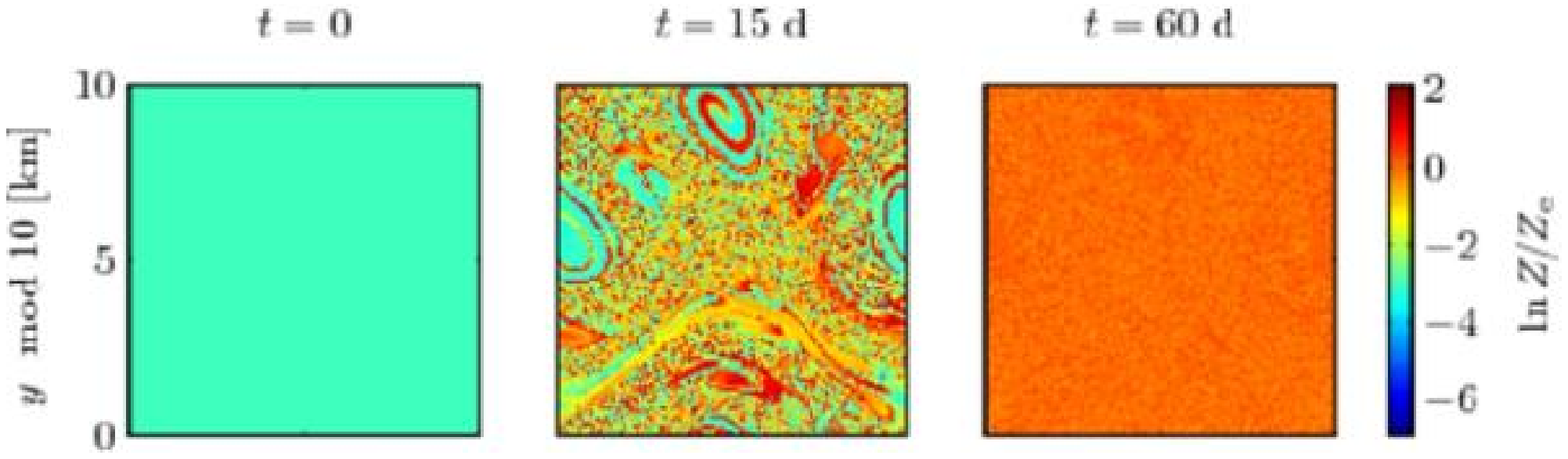}}%
\newline
\subfigure{\includegraphics[width=14cm]{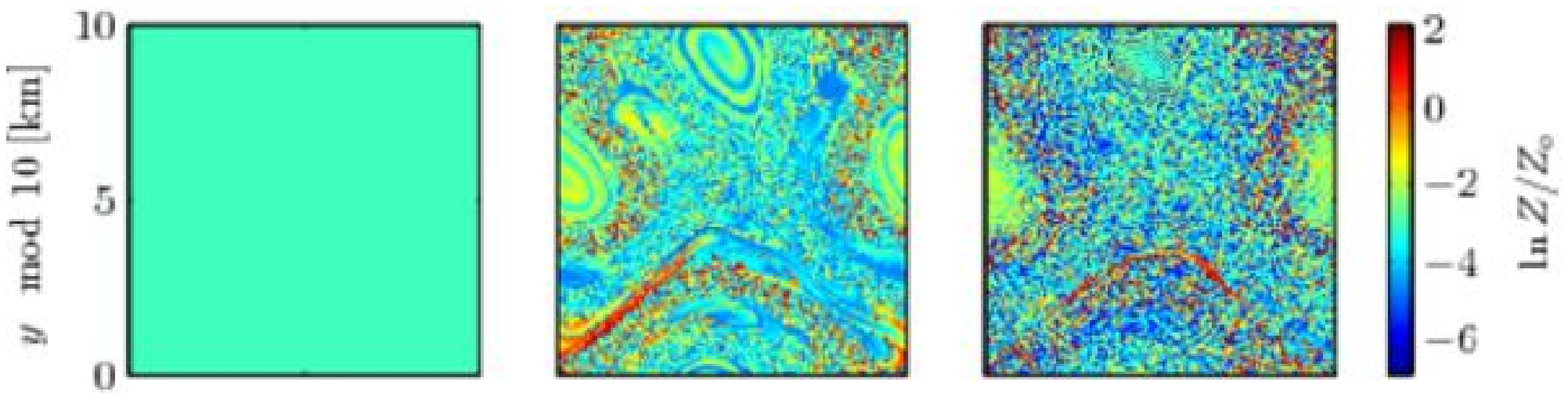}}\newline
\subfigure{\includegraphics[width=14cm]{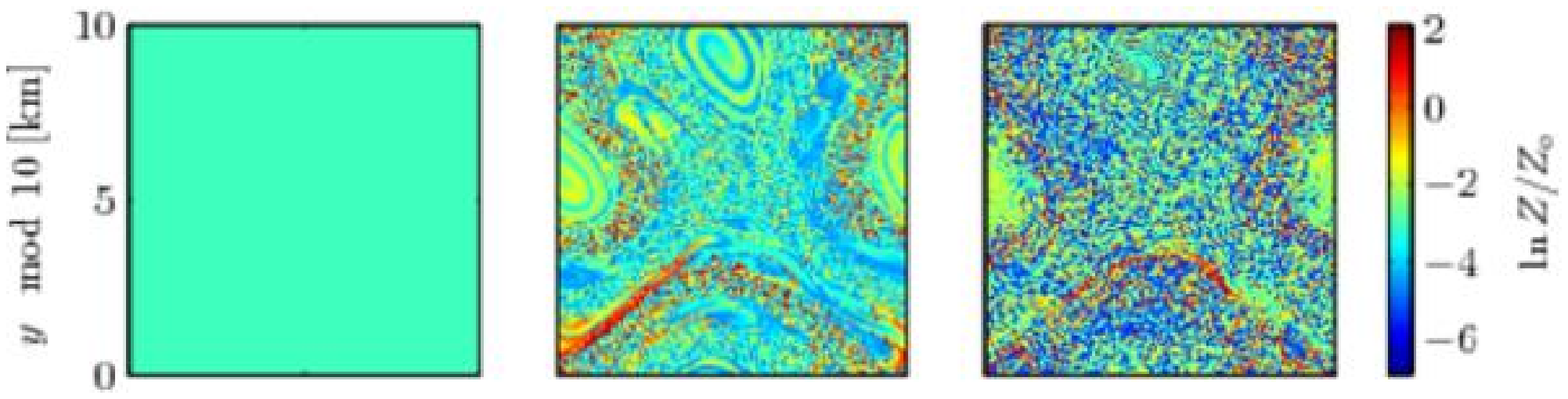}}\newline
\subfigure{\includegraphics[width=14cm]{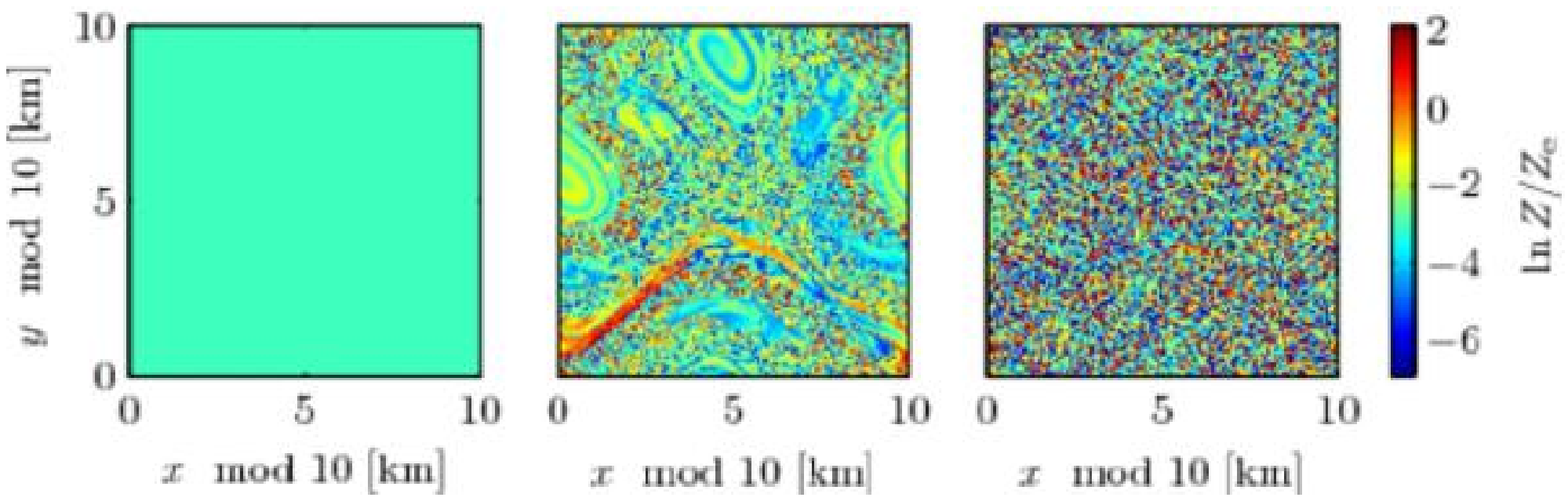}}
\caption{As in Fig. \protect\ref{P_t}, but for zooplankton.}
\label{Z_t}
\end{figure}

Figs. \ref{P_t} and \ref{Z_t} depict, respectively, possible evolution
scenarios for phytoplankton and zooplankton, which, while interacting
through predator--prey dynamics obeying (\ref{F}), are passively advected by
the same quasiperiodic tidal current involved in the construction of Figs. %
\ref{Poincare}--\ref{Tracer}. Predator--prey interaction parameters in the
top and mid-top row panels of Figs. \ref{P_t} and \ref{Z_t} are,
respectively, as in the left and right panels of Fig. \ref{PZ_1}.
Interaction parameters in the mid-bottom and bottom row panels of Figs. \ref%
{P_t} and \ref{Z_t} are, respectively, as in the left and right panels of
Fig. \ref{PZ_2}. Figs. \ref{P_t}--\ref{Z_t} were constructed by advecting $%
250000$ fluid particles, each one carrying phytoplankton and zooplankton
densities, distributed on regular grid defined on a doubly-periodic $10$ km $%
\times $ $10$ km domain of $\mathbb{R}^{2}$ as in Fig. \ref{Tracer}. The
spatial distributions of phytoplankton and zooplankton at any stage of the
evolution are obtained by triangle-based cubic interpolation onto the
regular grid based on Delaunay triangulation of the fluid particle positions.

At $t=0$ the phytoplankton distribution (Fig. \ref{P_t}, left column panels)
is assumed to be a Gaussian, while the zooplankton distribution (Fig. \ref%
{Z_t}, left column panels) is assumed to be uniform. This corresponds to a
sudden (nonuniform) availability of prey resources (phytoplankton blooming)
over the entire domain occupied by the predator. The locus in phase space of
the initial plankton distributions is indicated by blue dots in Fig. \ref%
{PZ_1}--\ref{PZ_2}.

Distributions of phytoplankton and zooplankton after $15$ d (resp., $60$ d)
are depicted in the middle (resp., right) column panels of Figs. \ref{P_t}
and \ref{Z_t}, respectively. The locus in phase space of these distributions
at $t=15$ d (resp., $t=60$ d) is indicated in Figs. \ref{PZ_1}--\ref{PZ_2}
with green (resp., red) dots. Note that $t=15$ d is still too early for the
regular autonomous predator--prey interactions in the left panel of Fig. \ref%
{PZ_1} to reach an asymptotically stable equilibrium, or for the regular
nonautonomous predator--prey interactions in the left panel of Fig. \ref%
{PZ_2} to settle on a periodic attracting set. Likewise, at $t=15$ d chaotic
predator--prey interactions in the right panel of Fig. \ref{PZ_2} produce
phytoplankton and zooplankton density values that cover a smaller portion of
the corresponding strange attractor than those attained at $t=60$ d. On the
other hand, approximately $15$ d must elapse for the autonomous
predator--prey interactions in the left panel of Fig. \ref{PZ_1} to reach a
limit-cycle oscillation.

\subsubsection{The Early Evolution Stage}

At $t=15$ d phytoplankton (Fig. \ref{P_t}, middle column panels) and
zooplankton (Fig. \ref{Z_t}, middle column panels) show structures in their
distributions that are very similar to those in the distribution of a
passive tracer (Fig. \ref{Tracer}, middle panel), which are an early
manifestation of Lagrangian mixed phase-space dynamics (Fig. \ref{Poincare},
right panel). Remarkably, although the estimated predictability horizon for
the chaotic predator--prey interactions is shorter than $15$ d, Lagrangian
mixed phase-space dynamics shows up very clearly in the plankton
distributions. The reason for this is that sensitivity to initial conditions
do not manifest at the same time for all orbits emanating from the latter in
the phase space for predator--prey interactions. A careful examination of
the mid panel in the bottom row of Figs. \ref{P_t}--\ref{Z_t}, however,
reveals small regions of very irregularly distributed plankton interspersed
within more uniformly distributed plankton, which are not a result of
chaotic stirring but rather of chaotic predator--prey interactions. Other
regions of very irregularly distributed plankton resulting from chaotic
interactions may lie hidden within the chaotically stirred plankton regions.
Clearly, for these regions to get exposed the timescale for substantial
development of chaotic stirring must be much larger than the predictability
horizon of chaotic predator-prey interactions.

\subsubsection{The Late Evolution Stage}

At $t=60$ d the regular predator--prey interactions in the left panel of
Fig. \ref{PZ_1} have almost reached an asymptotically stable equilibrium.
Accordingly, the right panel in the top row of Figs. \ref{P_t} and \ref{Z_t}
show very uniformly distributed phytoplankton and zooplankton
concentrations, respectively.

At the same instant $t=60$ d, regions of vigorously stirred phytoplankton
and zooplankton concentrations along with less stirred regions in the form
of filaments and spiral-like vortices develop when regular predator--prey
interactions autonomously cycle (Figs. \ref{P_t}--\ref{Z_t}, right panel in
the mid-top row) or nonautonomously behave on a $1:18$ subharmonic attractor
(Figs. \ref{P_t}--\ref{Z_t}, right panel in the mid-bottom row). These
features resemble very closely those of the passive tracer distribution at $%
t=60$ d in the right panel of Fig. \ref{Tracer}, which are a manifestation
of the mixed phase-space topology that characterizes Lagrangian dynamics.
However, note that for predator--prey interactions lying on the $1:18$
subharmonic attractor phytoplankton and zooplankton concentrations can take
at most 18 possible discrete asymptotic values (circles in the left panel of
Fig. \ref{PZ_2}). Consequently, at sufficiently long time these interactions
will lead to sharper color contrasts in the plankton distributions than
autonomously cycling interactions, for which plankton distributions take
continuous distributions.

Finally, $t=60$ d is far beyond the predictability horizon estimated for the
chaotic predator--prey interactions in the right panel of Fig. \ref{PZ_2}.
Correspondingly, the distributions of phytoplankton and zooplankton are
nearly completely mixed (Figs. \ref{P_t}--\ref{Z_t}, right panel in the
bottom row), which is a result of the exponential divergence in time of
initially nearby phytoplankton and zooplankton states rather than of chaotic
stirring.

\subsubsection{Remarks on Other Interaction Scenarios}

We close this section briefly discussing (but not showing) the results for
the two relevant intermediate nonautonomous biological interactions
scenarios mentioned earlier. These are nonautonomous interactions that take
place on a quasiperiodic cycle and nonautonomous interactions lying on
multiple periodic or quasiperiodic attractors. The quasiperiodically cycling
interaction case is not different than autonomously cycling interactions as
these are insensitive to initial conditions. The multiple attractor case, in
turn, is similar to the chaotic interaction case at long time because the
regular (periodic or quasiperiodic) attracting set reached by phytoplankton
and zooplankton densities is sensitive to their initial values.

\section{Discussion}

Suppose that random velocity field fluctuations of the form $\mathbf{u}%
^{\prime }=\sqrt{2D}(\eta _{1}(t),\allowbreak \eta _{2}(t))$ are
superimposed on $\mathbf{u}=(\partial _{y}\psi ,-\partial _{x}\psi )$, e.g.,
with $\psi $ defined by (\ref{psi}). Here $D$ is a constant and $\{\eta
_{i}\}$ are Gaussian-white-noise random variables satisfying $\left\langle
\eta _{i}\right\rangle =0$ and $\langle \eta _{i}(t)\eta _{j}(\tau )\rangle
=\delta _{ij}\delta (t-\tau )$, where the angle brackets denote ensemble
average. The expected distribution of a passive tracer, $\left\langle \theta
\right\rangle $, satisfies \citep[cf.,
e.g.,][]{Bazzani-etal-94} the Fokker--Plank (advection--diffusion) equation $%
\partial _{t}\left\langle \theta \right\rangle +[\left\langle \theta
\right\rangle ,\psi ]=D\Delta \left\langle \theta \right\rangle $ with
Fickian diffusivity constant $D$, where $\Delta $ is the Laplacian operator
in $\mathbb{R}^{2}$. Because $\mathbf{u}^{\prime }$ is a distribution rather
than a regular function, the resulting time-dependent vector field, $\mathbf{%
u}+\mathbf{u}^{\prime }$, is no longer Hamiltonian---at least in a strict
sense. Accordingly, KAM tori become permeable to transport and the resonance
islands turn leaky. In the long run, the mixed phase-space structures erode
completely, and consequently also the spatial patterns in tracer
distributions.

Standard ecological theories of pattern formation
\citep[cf.,
e.g.,][]{Murray-88} would approach the plankton distribution problem by
replacing (\ref{AT}) with diffusion equations coupled by predator--prey
dynamics, which is equivalent to considering (\ref{AT}) with a
Gaussian-white-noise random velocity field. Clearly, this approach cannot
result in a correct description of plankton advection by an oscillatory
velocity field such as the one considered here because the random velocity
field cannot produce Lagrangian mixed phase-space dynamics. Remarkably, it
has not been until very recently \citep[][]{Abraham-98} that standard
ecological theories have been explicitly recognized as being incapable of
providing an appropriate description of the formation of spatial patterns in
the distribution of plankton advected by ocean currents. In %
\citep[][]{Abraham-98} advection equations like (\ref{AT}) are considered
where the biological coupling is similar to the one considered here, except
that the carrying capacity is set to continually relax toward a spatially
nonuniform prescribed function (parameters are chosen so that the biological
interactions always contain an asymptotically stable equilibrium). The
velocity field in \citep[][]{Abraham-98} is a kinematic model for mesoscale
turbulence, which is envisioned as a superposition of randomly distributed
Gaussian eddies; generation of plankton patchiness is mainly attributed to
stirring by this velocity field.

The traditional approach to plankton advection by turbulent ocean currents
considers, instead of (\ref{AT}), a set of biologically-coupled
advection--diffusion equations with eddy diffusivities chosen several orders
of magnitude larger than molecular diffusivities. Clearly, this choice leads
to a complete erosion of any mixed phase-space signature underlying the
advection field at a much faster rate than molecular diffusion. More
sophisticated, higher-order Markovian stochastic models have been proposed %
\citep[][]{Berloff-McWilliams-02b} to account for the observed anomalous
(i.e., non-Brownian or non-random-walk) dispersion of passive tracers
advected by ocean currents, which we believe is more naturally explained by
Lagrangian mixed phase-space dynamics. Higher-order Markov models do not %
\citep[][]{Castronovo-Kramer-04} give a satisfactory description of the
anomalous dispersion behavior typically observed in the ocean. Other
stochastic models based on non-Gaussian (L\'{e}vy) statistics have been
proposed \citep[][]{Zaslavsky-94}, which result in diffusion-type equations
involving fractional partial derivatives. These so-called strange kinetic
theories, unlike the higher-order Markovian models, have the anomalous
dispersion behavior built in, but their practical utility is still not clear.

The present work provides a basis for an alternative approach to the problem
of pattern formation in plankton distributions in the presence of noisy
advection fields (i.e., large-scale advection fields with small-scale
environmental perturbations superimposed) or in turbulent open ocean
environments (e.g., mesoscale turbulence). The advection field in those
environments may be decomposed into some mean field plus noisy perturbations
(not necessarily small) due to environmental variability. The traditional
approach would treat the latter as stochastic. An approach in line with the
treatment given in this paper would consist in representing the noisy
perturbation field in Fourier space. Namely, through the superposition of
many harmonics (e.g., Rossby-wave-like modes in the case of mesoscale
turbulence) with random phases and amplitudes selected from some modeled or
observed wavenumber spectrum (e.g., plausibly following a power law in
mesoscale turbulence). For a given random phase realization, the result is a
velocity field with a streamfunction that in practice is of the
time-quasiperiodic form (\ref{xy}b). This allows for a quasi-deterministic
description wherein Lagrangian mixed phase-space dynamics plays a central
role.

\section{Summary and Conclusions}

In this paper we have investigated the effect Lagrangian dynamics on the
formation of spatial patterns in the distribution of plankton near the
surface of the ocean. Advection is assumed to be provided by a
divergence-free two-dimensional oscillatory velocity field. Unlike the
standard chaotic advection paradigm, in which case the oscillation of the
velocity field is periodic, the oscillation is here assumed to be
quasiperiodic, i.e., in the form of a superposition of periodic functions
with incommensurate frequencies. We analyzed in some detail the
two-frequency case, which is suitable for describing tidal motion in shallow
seas dominated by a predominantly semidiurnal regime including the
spring-to-neap cycle that results from superposing the semilunar and
semisolar tidal constituents. The results, however, apply also when many
more frequencies are considered, as would be required to model more
complicated (turbulent) advection fields. The evolution of phytoplankton and
zooplankton is assumed to be controlled by a set of Liouville (advection)
equations coupled by deterministic, spatially homogeneous predator--prey
dynamics of the standard Rosenzweig--MacArthur class. In addition to
autonomous interactions, we considered nonautonomous interactions by letting
the food availability to prey cycle diurnally.

The Lagrangian motion is described by a nonautonomous Hamiltonian system
with one degree of freedom. The phase space of this motion, which is the
physical space for advection, exhibits a mixed topological structure. This
includes regions of chaotic motion, possibly separated by KAM tori
(transport barriers), and populated by resonance islands (fluid particle
traps). The distribution of a passive tracer is thus seen to include
vigorously stirred regions and less stirred patches in the form of filaments
and spiral-like vortices identifiable with the features that comprise the
mixed phase space. These spatial patterns are subject to a quasiperiodic
vibration with basic frequencies given by the frequencies of the
time-quasiperiodic velocity field. In the particular case analyzed here, the
vibration is in synchrony with the assumed quasiperiodic tidal ebb and flood
surpluses. This contrasts with standard chaotic advection in which case such
a vibration is periodic.

The extent to which phytoplankton and zooplankton distributions are
determined by the Lagrangian mixed phase-space dynamics was shown to depend
on the type of predator--prey interactions taking place. We considered in
detail four possible interaction types: three regular (autonomous
stable-spiraling, autonomous cycling, and nonautonomous lying on a periodic
attractor) and one complex (nonautonomous taking place on a strange
attractor). For autonomously cycling interactions and nonautonomous
interactions lying on a periodic attractor, Lagrangian mixed phase-space
dynamics leads to the formation of spatial patterns (patchiness) in the
distribution of phytoplankton and zooplankton at all stages of the
evolution. For autonomous interactions including an asymptotically stable
equilibrium, phytoplankton and zooplankton distributions are determined by
the mixed phase-space structures during the early stages of the evolution,
before complete homogenization is reached. When the population dynamics is
chaotic, the influence of Lagrangian mixed phase-space dynamics is limited
to a time interval that is not too long compared to the predictability
horizon. Two further relevant nonautonomous biological interactions
scenarios were also discussed in the paper. These are nonautonomous
interactions taking place on a quasiperiodic cycle and lying on multiple
periodic or quasiperiodic attractors. In terms of plankton distributions,
the former is not very different than autonomously cycling interactions,
whereas the latter is similar to the long-term chaotic interaction case
because the regular (periodic or quasiperiodic) attracting set reached by an
orbit is sensitive to the initial conditions.

\noindent \textbf{Acknowledgements}

We thank I.~I. Rypina and I.~A. Udovydchenkov for the benefit of many
helpful discussions on Hamiltonian dynamics. This work has been supported by
NSF (USA) grant CMG-0417425 and PARADIGM NSF/ONR-NOPP (USA) grant
N000014-02-1-0370.

\bibliographystyle{elsart}

\vfill

\end{document}